\documentclass[aps,prx,twocolumn,groupedaddress,amsmath,epsfig,amssymb,eqsecnum]{revtex4}
\usepackage{makecell}
\usepackage{bbm}
\usepackage{mathrsfs}
\usepackage{amsfonts}
\usepackage{color}
\usepackage{graphicx}
\newcommand{\dbar}{d\hspace*{-0.08em}\bar{}\hspace*{0.1em}}
\begin{document}
\newcommand{\Rv}{{\vec {R}}} 
\newcommand{\rv}{{\vec r}}
\newcommand{\tv}{{\vec t}}
\newcommand{\av}{\boldsymbol a}
\newcommand{\fv}{{\boldsymbol f}}
\newcommand{\mv}{{\boldsymbol m}}
\newcommand{\nv}{{\boldsymbol n}}
\newcommand{\hv}{{\boldsymbol h}}
\newcommand{\xv}{{\boldsymbol x}}
\newcommand{\sv}{{\boldsymbol s}}
\newcommand{\xp}{\vec{x}_{\perp}}
\newcommand{\pp}{\vec{p}_{\perp}}
\newcommand{\zv}{{\boldsymbol z}}
\newcommand{\uv}{{\vec u}}
\newcommand{\Av}{{\boldsymbol A}}
\newcommand{\Xv}{{\boldsymbol X}}
\newcommand{\Yv}{{\boldsymbol Y}}
\newcommand{\Pv}{{\boldsymbol P}}
\newcommand{\Qv}{{\boldsymbol Q}}
\newcommand{\Hv}{{\boldsymbol H}}
\newcommand{\ur}{\vec{{\EuFrak u}}}
\newcommand{\cv}{{\vec c}}
\newcommand{\qv}{{\boldsymbol q}}
\newcommand{\pv}{{\boldsymbol p}}
\newcommand{\yv}{{\boldsymbol y}}
\newcommand{\vv}{{\boldsymbol v}}
\newcommand{\kv}{{\vec k}}
\newcommand{\phiv}{{\boldsymbol \phi}}
\newcommand{\etav}{{\boldsymbol \eta}}
\newcommand{\Tr}{{\rm Tr}}
\newcommand{\px}{{\partial_x}}
\newcommand{\py}{{\partial_y}}
\newcommand{\ppi}{{\partial_i}}
\newcommand{\ppj}{{\partial_j}}
\newcommand{\ch}{{\hat{c}}}
\newcommand{\eh}{{\hat{e}}}
\newcommand{\xh}{{\hat{x}}}
\newcommand{\yh}{{\hat{y}}}
\newcommand{\zh}{{\hat{z}}}
\newcommand{\vh}{{\hat{v}}}
\newcommand{\qh}{{\hat{q}}}
\newcommand{\kh}{{\hat{k}}}
\newcommand{\llm}{{\boldsymbol \lambda}}
\newcommand{\Am}{{\boldsymbol{A}}}
\newcommand{\Qm}{{\boldsymbol{Q}}}
\newcommand{\Rm}{{\boldsymbol R}}
\newcommand{\Lm}{{\boldsymbol L}}
\newcommand{\Km}{{\boldsymbol K}}
\newcommand{\Jm}{{\boldsymbol J}}
\newcommand{\Tm}{{\boldsymbol T}}
\newcommand{\Bm}{{\boldsymbol B}}
\newcommand{\Dm}{{\boldsymbol D}}
\newcommand{\Cm}{{\boldsymbol C}}
\newcommand{\Em}{{\boldsymbol E}}
\newcommand{\Mm}{{\mathcal M}}
\newcommand{\Wm}{{\boldsymbol W}}
\newcommand{\Fm}{{\boldsymbol F}}
\newcommand{\Gm}{{\boldsymbol G}}
\newcommand{\Imm}{{\boldsymbol I}}
\newcommand{\sm}{{\boldsymbol s}}
\newcommand{\gammam}{{\boldsymbol \Gamma}}
\newcommand{\chim}{{\boldsymbol \chi}}
\newcommand{\be}{\begin{equation}}
\newcommand{\ee}{\end{equation}}
\newcommand{\ba}{\begin{eqnarray}}
\newcommand{\ea}{\end{eqnarray}}
\newcommand{\RNum}[1]{\uppercase\expandafter{\romannumeral #1\relax}}
\newcommand{\ddelta}{\boldsymbol{\delta}}
\newcommand{\pdf}{\mathcal{P}}
\newcommand{\U}{\mathcal{U}}
\newcommand{\LFP}{\mathcal{L}_{\rm FP}}

\title{Microscopic Dynamical Entropy\\ II: Statistical and Stochastic Thermodynamics of Hamiltonian Systems}
\author{Mingnan Ding}
\email{dmnphy0@gmail.com}
\affiliation{Wilczek Quantum Center, School of Physics and Astronomy, Shanghai Jiao Tong University, Shanghai 200240, China}
\affiliation{DAMTP, Centre for Mathematical Sciences, University of Cambridge, Wilberforce Road, Cambridge CB3 0WA, United Kingdom}

\author{Michael E. Cates}
\affiliation{DAMTP, Centre for Mathematical Sciences, University of Cambridge, Wilberforce Road, Cambridge CB3 0WA, United Kingdom}

\date{\today} 

\begin{abstract}

The Microscopic Dynamical Entropy (MDE) introduced in Ref.~\cite{Ding2026} describes irreversible relaxation of selected variables $\xv$ within a finite closed Hamiltonian system of fixed total energy $E$. Here we extend the framework to systems with arbitrary distributions of the total energy and to cases where the Hamiltonian is time-dependent. The latter allows us to address work interactions with external agents or protocols -- which are crucial in both stochastic and classical thermodynamics -- as well as driven systems more generally. The central step is to construct, from the full microscopic description in terms of the selected variables $\xv$ and the unmonitored variables $\yv$, a reduced description in terms of the selected variables and the instantaneous total energy $(\xv,E)$. The unmonitored degrees of freedom $\yv \in$ Y enter this description through their conditional phase-space volume (CPV) $\Omega_Y(\xv,E,t)$.
This construction allows work and heat to be defined directly for a finite Hamiltonian system comprising a subsystem X coupled to a heat reservoir Y, including cases where Y is so small that its temperature is not a well-defined thermodynamic quantity. We demonstrate monotonic MDE growth (the macroscopic second law) under mixing and time-scale separation even for driven systems with arbitrary energy distributions.
At the trajectory level, we derive detailed and integral fluctuation relations for the MDE, generalising what is well known for the case of an unbounded heat bath. We verify these numerically in a driven few-particle Hamiltonian system, showing emergence of stochastic thermodynamics for collective coordinates (such as the centre-of-mass position), even in closed systems with only ten or twenty degrees of freedom in total. Physically, these fluctuation relations connect irreversibility to the change in the number of unmonitored microstates compatible with initial and final data for the observed coordinates. Together, our results provide a finite-system Hamiltonian foundation for the emergence of classical and stochastic thermodynamics and extend their applicability to surprisingly small heat baths.

\end{abstract}

\maketitle

\section{Introduction}

Understanding the microscopic origin of thermodynamic behavior remains a fundamental problem of statistical physics. Classical Hamiltonian dynamics is time-reversible and preserves phase-space volume via Liouville's theorem, yet macroscopic systems exhibit irreversible behavior described by the second law of thermodynamics~\cite{Clausius1854,Clausius1865,Carnot1872}. Traditional approaches connect these descriptions through statistical ensembles and coarse-graining, leading to the Gibbs and Boltzmann entropies in equilibrium~\cite{Boltzmann2012,Gibbs1902,Jaynes1965,Ehrenfest1959}. While highly successful, these formulations do not provide a microscopic framework for nonequilibrium thermodynamics in finite Hamiltonian systems~\cite{Lebowitz2005,Uffink2001,teVrugt2021,Pathria2011,Sethna2006}.
The full system, including the bath, evolves under deterministic Hamiltonian dynamics. Thermodynamic behavior must therefore be understandable directly at this level. This raises a central question: how to define an entropy, evolving under Hamiltonian dynamics, that captures irreversibility without introducing unnecessary or unjustified assumptions.

In Ref.~\cite{Ding2026} (hereafter, Paper I), we introduced a microscopic dynamical entropy (MDE) for a time-independent Hamiltonian system XY moving on an energy shell $E$, whose degrees of freedom are partitioned into monitored variables $\xv$ and unmonitored variables $\yv$ in subsystems X,Y $\in$ XY. The central object is the conditional phase-space volume (CPV) $\Omega_Y(\xv,E)$, which measures the volume of configurations $\yv$ compatible with given values of both $\xv$ and $E$.

Our MDE is determined by the geometry of the conditional phase space, rather than any {\em a priori} distribution or ensemble governing the monitored variables $\xv$. This reflects the defining principle of thermal entropy: it depends only on macroscopic constraints and discards microscopic bath details. The MDE realizes this principle microscopically through $\Omega_Y(\xv,E)$. In Paper I it was shown that, when the variables $\xv$ are slow and evolve autonomously at fixed $E_{XY}=E$, the MDE is non-decreasing under Hamiltonian dynamics, providing a microscopic formulation of the second law for relaxation processes. This contrasts with the full Gibbs entropy $S_{XY}^G$ of the XY system which is famously constant under Hamiltonian evolution (Liouville's theorem) and is thus not a credible MDE candidate~\cite{Ding2026}.

The present work establishes a natural reduced description of Hamiltonian dynamics in terms of the variables $(\xv,E)$, obtained from the full phase space $(\xv,\yv)$ by eliminating unmonitored degrees of freedom $\yv$. This induces a probabilistic structure on the reduced state space, governed by the conditional phase-space volume $\Omega_Y(\xv,E)$, and provides a foundation for defining nonequilibrium entropy without introducing stochastic assumptions. Crucially, and in contrast to Paper I, we allow $E$ to vary and thus can address work interactions (or equivalently, time-dependent Hamiltonians), alongside heat flow. Such interactions are of course fundamental both to classical and stochastic thermodynamics.

Within this newly extended formulation, the MDE is defined on the reduced state space $(\xv,E)$ and admits a general expression for arbitrary phase-space distributions. 
The construction is expressed entirely in terms of $(\xv,E)$, the marginal phase-space density $\rho_{XE}(\xv, E,t)$, the Hamiltonian $H_{XY}(t)$, and its conditional phase-space volume $\Omega_Y(\xv,E)$, without taking thermodynamic quantities such as temperature as {\em a priori} inputs; instead the latter emerge when interpreting heat exchange and other thermal properties. 

The resulting MDE separates into a structural contribution determined by $\Omega_Y(\xv,E)$ and a term describing deviations from equilibrium, showing how thermodynamic entropy connects to the underlying phase-space geometry. The same construction also yields a natural generalization of the Hamiltonian of mean force, previously defined only for a system in contact with a thermodynamically large heat bath~\cite{jarzynski2004,Campisi2009,Talkner2020,Strasberg2021}, to finite Hamiltonian systems. We allow the subsystem energy to change through both external driving and energy exchange with the remaining degrees of freedom.

At the dynamical level, we shall see that the trajectory-level MDE defines an entropy production for the reduced dynamics of $\xv$ variables in the presence of unmonitored $\yv$ degrees of freedom. The resulting fluctuation relation follows directly from micro-reversibility and Liouville's theorem~\cite{Evans1993,Gallavotti1995,Evans2002}, and recovers standard results such as the Crooks fluctuation theorem and the Jarzynski equality in the large-bath limit~\cite{Jarzynski1997,Crooks1999,Seifert2005,Jarzynski2011}. Thus the present work directly connects the MDE of Paper I to stochastic thermodynamics, by allowing arbitrary work protocols that change the (system+bath) energy $E_{XY}=E$. Stochastic thermodynamics deals extensively with such protocols for the case of a small system X and a large bath Y (see \cite{Seifert2005,Seifert2012}), whereas our work extends beyond the large-bath restriction.

Together with Paper I, the results presented here provide a unified framework in which probability, entropy, thermodynamic structure, and irreversibility emerge from the geometry of the conditional phase space encoded in $\Omega_Y(\xv,E)$ on the reduced state space $(\xv,E)$. Although our work does not solely address the case where Y is a heat bath, in that case the function $\Omega_Y(\xv,E)$ represents the minimal information about the bath that allows thermodynamics to be constructed from Hamiltonian mechanics. This applies equally to small systems X as to large ones and thus holds equally for both stochastic and classical thermodynamics.

The paper is organized as follows. Section~\ref{sec:mde-general-energy} introduces the MDE for general energy distributions, discusses its relation to the Gibbs entropy of the total system, and develops its dynamical properties, including monotonicity (under the autonomy assumption discussed in Paper I) and the associated fluctuation relation. Section~\ref{sec:thermo} develops the thermodynamic structure implied by the MDE and shows its reduction to standard thermodynamics in the large-bath limit. Section~\ref{sec:example} presents an explicit driven-particle example in which the change in MDE can be computed analytically at trajectory level. This example verifies the trajectory-level fluctuation relation (in a system with a surprisingly small bath). Finally, Sec.~\ref{sec:conclusion} summarizes the results and discusses their implications.

\section{Microscopic dynamical entropy for general energy distributions}
\label{sec:mde-general-energy}

In this Section we construct the MDE formalism for systems XY based on the variables $(\xv,E)$, where $\xv$ denotes the degrees of freedom of the studied subsystem X and $E$ the total energy which includes that of the ``bath'' Y. 
The inclusion of $E$ as a dynamical variable (taken constant in Paper I) is essential to capture changes in energy associated with work interactions, or more general driving via a time dependent Hamiltonian $H_{XY}(t)$. The resulting representation of the full phase space in terms of $(\xv,E)$ remains well-defined for arbitrary, time-dependent phase-space distributions.

The extension of the MDE construct to variable $E$ allows a change of viewpoint, whereby the mapping from $(\xv,\yv)$ to $(\xv,E)$ induces a probabilistic structure in which the unmonitored degrees of freedom $\yv$ are represented through the conditional phase-space volume $\Omega_Y(\xv,E)$. In this representation, the state of the system X and its bath Y is described by the energy-resolved distribution $\rho_{XE}(\xv,E,t)$, and the MDE becomes a functional of this distribution. This provides a unified framework in which both equilibrium and nonequilibrium thermodynamics are determined by the geometry of phase space.

\subsection{MDE on a single energy shell}
\label{sec:mde-shell}

From the above viewpoint, we first briefly recall the MDE for a system XY confined to a fixed energy shell $E_{XY}=E$~\cite{Ding2026}.
Consider a time-independent Hamiltonian $H_{XY}(\xv,\yv)$, where $\xv$ denotes a chosen subset of canonical variables (coordinates and momenta) and $\yv$ represents the remaining degrees of freedom. The decomposition (X,Y) does not necessarily correspond to two physically separable subsystems, although Y will below play the role of an effective bath for X.

For any given values $(\xv,E)$ the conditional phase-space volume (CPV) is defined as
\ba
\Omega_Y(\xv,E)
\equiv \int d\yv \, \delta\!\left(E-H_{XY}(\xv,\yv)\right).
\label{OmegaY-shell}
\ea

The total phase-space volume of a given energy shell is
\ba
Z(E) = \int d\xv\, \Omega_Y(\xv,E).
\label{Z-shell}
\ea
Its logarithm is the Boltzmann entropy (we set $k_B= 1$ throughout).
The CPV defines a reference distribution of $\xv$ on the given energy shell,
\ba
\rho_{X|E}^{\rm ref}(\xv,E)
= \frac{\Omega_Y(\xv,E)}{Z(E)}.
\label{rhoXeq-shell}
\ea
Under appropriate dynamical conditions, such as mixing dynamics for the XY system~\cite{Arnold1968,Walters1982,Cornfeld1994}, this is the equilibrium distribution of $\xv$ on the energy shell (see Paper I).

The MDE is defined as 
\ba
S^{\rm md}_{XY|E}(E,t)
&=& S^G_{X|E}[\rho_{X|E}(\xv, E, t)]
\label{mde-shell-def}\\
&&+ \int d\xv \rho_{X|E}(\xv,E,t)\log\Omega_Y(\xv,E,t), \nonumber
\ea
where the first term is the Gibbs entropy of the marginal distribution $\rho_{X|E}(\xv,E,t)$ for subsystem X given that $E_{XY}=E$:
\ba
S_{X|E}^G[\rho_{X|E}(\xv, E, t)]
&=& - \int d\xv \rho_{X|E}
\log\rho_{X|E}.
\label{SGX-shell}
\ea

The choice of selected variables X is central to any practical use of the MDE framework; it defines the scope of the reduced description, via $\rho_{X|E}(\xv, E, t)$, of the full Hamiltonian dynamics of XY.  In particular, monotonicity (in time) of the MDE is not guaranteed for arbitrary choices of X, but holds when the dynamics of the reduced variables is autonomous in the sense explained in Paper I. This typically arises under time-scale separation between X and Y variables. 

We finally note that, as discussed in Paper I, the choice of the selected variables $\xv$ is not restricted to canonical coordinates. More generally, if $\Gamma$ denotes the full microscopic phase-space point and $\xv=X(\Gamma)$ is an arbitrary selected variable, the CPV should be written as
\ba
\Omega_Y(\xv,E)
&=&
\int d\Gamma\,
\delta\left(\xv-X(\Gamma)\right)
\delta\left(E-H(\Gamma)\right).
\label{OmegaY-general-X}
\ea
The form in Eq.~(\ref{OmegaY-shell}) is the adapted-coordinate version of this more general definition, obtained when the canonical phase-space variables can be written as $\Gamma=(\xv,\yv)$. In the present paper we use Eq.~(\ref{OmegaY-shell}) for notational simplicity. 

\subsection{MDE in full phase-space}
\label{sec:mde-gibbs}

Extending the definition of MDE to systems with non-trivial energy distribution is important for two reasons. First, when the total Hamiltonian $H_{XY}(\xv,\yv,t)$ depends explicitly on time, the total energy $E$ is not conserved, so a formulation based on one fixed shell is not appropriate. Such a time-dependent $H_{XY}$ arises whenever a system X, in thermal contact with its bath Y, is also coupled to an external machine able to impart or extract energy from the XY system as work. Such work interactions are of course pivotal to classical thermodynamics and the study of time-dependent Hamiltonian protocols is central to stochastic thermodynamics as well~\cite{Seifert2005,Seifert2012}. Second, even for a time-independent Hamiltonian, it is useful to have a form applicable to a general phase-space density $\rho_{XY}(\xv,\yv,t)$, which may have support on many energy shells simultaneously. However, since the probability density for $E$ is time-independent, the generalization of the MDE is in that case straightforward.

In either case we first resolve the full distribution with respect to the instantaneous total energy:
\ba
\rho_{XE}(\xv,E,t)=\int d\yv\, \rho_{XY}(\xv,\yv,t)\delta\left(E-H_{XY}(\xv,\yv,t)\right). \nonumber\\
\label{rho-XE-def}
\ea
The corresponding marginal distribution of energy is
\ba
\rho_E(E,t)=\int d\xv\, \rho_{XE}(\xv,E,t),
\label{rho-E-def}
\ea
and the conditional distribution of $\xv$ at total energy $E_{XY}=E$ is
\ba
\rho_{X|E}(\xv,E,t)=\frac{\rho_{XE}(\xv,E,t)}{\rho_E(E,t)}.
\label{rho-X-given-E}
\ea
For each energy shell the CPV of the bath variables Y is
\ba
\Omega_Y(\xv,E,t)=\int d\yv\, \delta\left(E-H_{XY}(\xv,\yv,t)\right).
\label{OmegaY-general}
\ea

From (\ref{rho-XE-def}--\ref{OmegaY-general}) we define a reference distribution
\ba
\rho_X^{\rm ref}(\xv,E,t) \equiv \frac{\Omega_Y(\xv,E,t) }{ Z(\xv, E, t)}
\equiv \frac{ \Omega_Y(\xv,E,t) }{ \int \Omega_Y(\xv,E,t) d\xv}.
\ea
For systems that are ergodic (or mixing) on the energy shell, this reference distribution coincides with the equilibrium distribution. Such systems constitute the main focus of statistical mechanics.

To define the MDE we construct from 
the full microscopic
phase-space density $\rho_{XY}(\xv,\yv,t)$ the following `flattened' distribution
\ba
\tilde\rho_{XY}(\xv,\yv,t)
&=& \int dE \rho_{XE}(\xv,E,t)
\frac{\delta\left(E-H_{XY}(\xv,\yv,t)\right)}
{\Omega_Y(\xv,E,t)} \nonumber\\
&=&
\frac{
\rho_{XE}\left(\xv,H_{XY}(\xv,\yv,t),t\right)
}{
\Omega_Y\left(\xv,H_{XY}(\xv,\yv,t),t\right)
}.
\label{rho-tilde-explicit}
\ea
(The MDE of Paper I is recovered when $\rho_{XE}$ has support on a single $E$ value.) The flattened distribution retains all information contained in $\rho_{XE}(\xv,E,t)$ while discarding the internal details of the microstates $\yv$ compatible with any given $(\xv,E)$. More precisely, for each $(\xv,E)$,  $\tilde\rho_{XY}$ replaces the conditional probability density $\rho_{Y|X,E}$ with a $\yv$-independent density of the same total probability mass. The MDE is then the Gibbs entropy of the flattened distribution, as elaborated in \eqref{mde-def-gibbs} below and, for the constant energy case, fully discussed in Paper I.

A crucial feature of the present construction is that the reduction from $(\xv,\yv)$ to $(\xv,E)$ constitutes a {\em non-destructive} coarse-graining~\cite{Ding2026}. By this we mean that the distribution $\rho_{XE}(t)$ remains entirely determined by the underlying Hamiltonian dynamics; we make no modification of the microscopic evolution. That is, the full distribution $\rho_{XY}(t)$ always evolves according to exact Hamiltonian dynamics, and $\rho_{XE}(t)$ is obtained from it through Eq.~(\ref{rho-XE-def}). 
This is fundamentally different from conventional phase-cell coarse-graining~\cite{Ehrenfest1959}, where the dynamics itself is altered so as to ensure monotonic entropy increase. By preserving the exact microscopic reversibility and Liouville structure, the MDE avoids conceptual difficulties such as reversibility paradoxes exemplified by spin-echo protocols~\cite{Ridderbos1998}.
As discussed further in Paper I, this feature of the MDE ensures that irreversibility arises solely from the discarding of information about the heat bath Y, rather than by replacing the Hamiltonian dynamics of the  XY system with some altered, inexact dynamics.

As promised we now define the MDE as the Gibbs entropy of the non-destructive coarse-grained distribution:
\ba
S^{\rm md}_{XY}[\rho_{XY}]
&\equiv&
S^G_{XY}[\tilde\rho_{XY}]\nonumber \\
&=& -\int d\xv d\yv \tilde\rho_{XY}\log \tilde\rho_{XY}.
\label{mde-def-gibbs}
\ea
Using Eq.~(\ref{rho-tilde-explicit}) we obtain
\ba
S^{\rm md}_{XY}(t)
&=& -\int d\xv d\yv \tilde\rho_{XY}
\log \rho_{XE}(\xv,H_{XY},t)
\nonumber \\
&+& \int d\xv d\yv \tilde\rho_{XY}
\log \Omega_Y(\xv,H_{XY},t).\label{two-one-six}
\ea

We now note that for an arbitrary function $f(E)$ we have from \eqref{rho-tilde-explicit} the identity
\ba
\int d\yv f(H_{XY})
&=&\int d\yv\int dE f(E)\delta\left(E-H_{XY}\right) \nonumber\\
&=&  \int dE f(E)\Omega_Y(\xv,E,t).
\label{energy-shell-identity}
\ea
Applying this to \eqref{two-one-six} gives 
\ba
S^{\rm md}_{XY}(t)
&=& -\int d\xv dE \rho_{XE}(\xv,E,t)\log \rho_{XE}(\xv,E,t)\nonumber \\
&& + \int d\xv dE \rho_{XE}(\xv,E,t)\log \Omega_Y(\xv,E,t). \nonumber\\
&=&  S_E^G[\rho_E] + \bigg\langle S_X^G[\rho_{X|E}] \nonumber\\
&& + \int d\xv \rho_{X|E}(\xv,E,t)\log \Omega_Y(\xv,E,t)
\bigg\rangle_E \nonumber\\
& = & S_E^G[\rho_E] + \bigg\langle S^{\rm md}_{XY|E}(E,t) \bigg \rangle_E
\label{mde-shell-decomposition}
\ea
where $\langle \rangle_E$ denotes average over $\rho_E$.
This result shows that the MDE naturally separates into the Gibbs entropy of the energy distribution plus the average of the microcanonical MDE (as defined in Paper I) associated with each energy shell. (The first term $S^G_E$ becomes singular
in the microcanonical limit as already discussed there.) 
Clearly the MDE is now a functional of $\rho_{XE}$, instead of $\rho_X$ as in Paper I. 

For time-independent Hamiltonians (only) the energy distribution $\rho_E(E)$ is conserved under Hamiltonian dynamics, so $S_E^G[\rho_E]$ is constant. Subtracting that constant, the full MDE is effectively just the average of the microcanonical one over the fixed energy distribution. This could easily have been guessed. However it is important, and not easily guessed, that \eqref{mde-shell-decomposition} continues to be equal to $S^G_{XY}[\tilde\rho_{XY}]$ in \eqref{rho-tilde-explicit}
even for time-dependent Hamiltonians $H_{XY}$. It is this, and not \eqref{mde-shell-decomposition} itself, that allows the quantitative correspondence between our MDE and the thermodynamic entropy to be maintained in the presence of work interactions.

It is important that $E$ denotes the total energy of the full $XY$ system. At finite coupling, separating this total energy into subsystem and bath contributions is generally ambiguous, especially for a finite total system. By using $(\xv,E)$ as the reduced variables, our MDE construction avoids this ambiguity: $\Omega_Y(\xv,E)$ is always defined directly from the full Hamiltonian and counts the Y-states compatible with the chosen $\xv$ and the total energy $E$. This distinction becomes important beyond weak coupling and will be further discussed in Sec.~\ref{sec:EX}.

\subsection{Alternative form of MDE}

Using the quantities defined above, the entropy contribution associated with each energy shell can be rewritten in a useful form. 
Starting from Eq.~(\ref{mde-shell-def}), we obtain
\ba
&& S_X^G[\rho_{X|E}] +
\int d\xv \rho_{X|E}(\xv,E,t)\log \Omega_Y(\xv, E,t) \nonumber \\
&=& \log Z(E,t) - D\left(
\rho_{X|E}(\xv,E,t) \middle\| \rho_X^{\rm ref}(\xv,E,t) \right), \nonumber
\ea
(with $Z(E,t)$ defined as in \eqref{Z-shell}) where
\ba
&&D\left(
\rho_{X|E}(\xv,E,t) \middle\| \rho_X^{\rm ref}(\xv,E,t) \right) \\
&=& \int d\xv  \rho_{X|E}(\xv,E,t) 
\log\frac{\rho_{X|E}(\xv,E,t)  }{ \rho_X^{\rm ref}(\xv,E,t)  }
\ea
denotes the Kullback--Leibler divergence between the actual and the reference distributions.
Substituting this relation into the definition \eqref{mde-def-gibbs} of the MDE yields 
\ba
S^{\rm md}_{XY}(t)
&=& S_E^G[\rho_E] + \int dE \rho_E(E,t)\log Z(E,t)
\label{mde-kl-form}\\
&-& \int dE \rho_E(E,t) D\left( \rho_{X|E}(\xv,E,t) \middle\| \rho_X^{\rm ref}(\xv,E,t)
\right).
\nonumber
\ea
This expression separates the MDE into three distinct contributions. The first term $S_E^G[\rho_E]$ is the Gibbs entropy of the energy distribution, which quantifies the uncertainty in the total energy of the system. Since work done by or on an external protocol is the sole cause of changes in the total energy $E$ of the XY system, the distribution $\rho_E$ encodes the statistics of work.

The second term $\int dE  \rho_E(E,t)\log Z(E,t)$ is determined by the total phase-space volume of each energy shell and reflects the geometric structure of the accessible microstates.
Together, these two terms depend only on the distribution over energy and the phase-space volume of each shell, and represent a structural (Boltzmann-type) contribution to the entropy, which indeed is the {\em only} contribution for a system in thermal equilibrium.

The last term measures the deviation of the actual phase-space distribution from the reference distribution on each energy shell. Its decrease under the dynamics characterizes relaxation toward this distribution and quantifies the part of the entropy production associated with equilibration on each energy shell.

Equation~\eqref{mde-kl-form} also shows that, although for a time-independent Hamiltonian the MDE contains the same dynamical information as the KL divergence, the latter alone cannot serve as the microscopic definition of a dynamical entropy in systems where the Hamiltonian is time-dependent. The KL relative entropy has appeared previously in information-theoretic and stochastic-thermodynamic formulations of nonequilibrium behavior \cite{Cover1999,Jarzynski2011,Seifert2012}. Our result does not invalidate such constructions. Rather, Eq.~\eqref{mde-kl-form} shows how the KL structure emerges directly from Hamiltonian dynamics through the conditional phase volume. More importantly, when the total energy itself is distributed, an additional contribution $S_E^G[\rho_E]$ appears naturally. This contribution plays no role on a fixed energy shell but becomes essential once work exchanges are allowed and the energy evolves dynamically.

\subsection{Autonomy condition and monotonic MDE}
\label{sec:2nd-law}

We now show that the MDE is non-decreasing when the dynamics of $(\xv,E)$ is autonomous. 
The argument follows closely that of Paper I, but now the MDE is a functional of the full reduced distribution $\rho_{XE}$ rather than the microcanonical $\rho_{X|E}$. 

Let $\rho_{XY}(\xv,\yv,t)$ denote the phase-space distribution of the full $XY$ system, and let $\tilde\rho_{XY}(t_0)$ be the flattened distribution defined in Eq.~(\ref{rho-tilde-explicit}). By construction,
\ba
S^{\rm md}_{XY}[\rho_{XY}(t_0)] = S^G_{XY}[\tilde\rho_{XY}(t_0)].
\ea
Evolving $\tilde\rho_{XY}(t_0)$ under Hamiltonian dynamics to time $t_1$ gives an auxiliary distribution $\rho^*_{XY}(t_1)$. By Liouville's theorem,
\ba
S^G_{XY}[\rho^*_{XY}(t_1)] = S^G_{XY}[\tilde\rho_{XY}(t_0)].
\ea
The evolved distribution $\rho^*_{XY}(t_1)$ will not in general be conditionally uniform in Y at fixed $(\xv,E)$. Reapplying the flattening procedure gives
\ba
&&S^{\rm md}_{XY}[\rho^*_{XY}(t_1)]
\equiv S^G_{XY}[\tilde\rho^*_{XY}(t_1)]
\nonumber\\
&\ge& S^G_{XY}[\rho^*_{XY}(t_1)]
= S^{\rm md}_{XY}[\rho_{XY}(t_0)],
\label{coarse}
\ea
where the inequality follows because flattening the conditional distribution of Y at fixed $(\xv,E)$ cannot decrease the Gibbs entropy.

We now assume that the true distribution $\rho_{XY}(t_1)$ and the auxiliary distribution $\rho^*_{XY}(t_1)$ have almost the same reduced distribution:
\ba
\rho_{XE}(t_1) = \rho^*_{XE}(t_1) + O(\epsilon).
\label{endpoint-cond}
\ea
Here $\epsilon$ is a small parameter that is system specific but is usually a time-scale separation parameter; see Paper I for a further discussion.
Since the MDE depends only on $\rho_{XE}$ the MDE then satisfies
\ba
S^{\rm md}_{XY}[\rho_{XY}(t_1)]
\ge
S^{\rm md}_{XY}[\rho_{XY}(t_0)]
+O(\epsilon).
\label{SecondL}
\ea
Equation~(\ref{endpoint-cond}), with the $O(\epsilon)$ term neglected, represents an {\em autonomy condition} for the reduced variables $(\xv,E)$: their distribution at $t_1$ is determined by that at $t_0$, and not by the details of Y. If this holds for all $t_1>t_0$, the MDE evolves monotonically in time.

\subsection{Trajectory-level MDE and fluctuation relation}
\label{sec:MDEft}

So far we have formulated the MDE, $S^{\rm md}_{XY}$, as a functional of the time-dependent phase-space density $\rho_{XY}(t)$. However, \eqref{mde-def-gibbs} shows this to be the average over that density (referred to loosely below as the `ensemble average') of the following trajectory-level quantity: 
\ba
s^{\rm md}(\xv,E,t)
\equiv
-\log \rho_{XE}(\xv,E,t)
+\log \Omega_Y(\xv,E,t),\nonumber \\
\label{sMDE-def}
\ea
This is well defined for any trajectory and energy-interaction protocol specified within the reduced dynamical variable set $(\xv(t),E(t))$; as just stated, its `ensemble average' gives the MDE. Although $s^{\rm md}$ depends on the ensemble distribution $\rho_{XE}$, it is evaluated along individual trajectories $(\xv(t),E(t))$ . We take \eqref{sMDE-def} to define a trajectory level MDE, and show below how this recovers the expected properties of such an entropy -- specifically, the fluctuation theorem(s) of stochastic thermodynamics. 

We consider Hamiltonian evolution in the full phase space $(\xv,\yv)$ between $t=0$ and $t=\tau$, and restrict the XY Hamiltonian to be even in momenta. 
Since the mapping from $(\xv,\yv)$ to $(\xv,E)$ is many-to-one, the induced dynamics of $(\xv,E)$ is probabilistic. To define this reduced dynamics, one must specify a measure over the unobserved variables $\yv$ consistent with a given $(\xv,E)$.

Our proof of the fluctuation theorem takes the initial microscopic ensemble to be the flattened distribution associated with the chosen initial reduced state \footnote{Or, more precisely, to be equivalent to the flattened distribution for the purposes of calculating entropies.}. Importantly, this is not an equilibrium assumption on the {\em reduced} variables: the distribution $\rho_{XE}(\xv,E,0)$ remains arbitrary. However we do assume that, at $t=0$, the unmonitored variables $\yv$ are conditionally equilibrated subject to $(\xv,E)$. As discussed in Sec.~\ref{sec:2nd-law}, this is most easily justified via a time-scale separation between a fast bath Y and slower selected variables X -- an assumption generally left implicit in stochastic thermodynamics.

With this assumption, we have
\ba
\tilde\rho_{XY}(\xv,\yv,0)
= \int dE
\rho_{XE}(\xv,E,0)
\frac{\delta\left(E-H_{XY}(\xv,\yv,0)\right)}
{\Omega_Y(\xv,E,0)}. \nonumber\\
\label{rho-tilde-main}
\ea
The derivation below relies on the same conditional-equilibration assumption for the unmonitored variables Y that underlies the construction of the flattened distribution in Eq.~(\ref{rho-tilde-main}).

The probability $\mathcal P_F$ that the system reaches $(\xv',E')$ at time $\tau$ can then be written as
\ba
\mathcal P_F(\xv',E')
= \int d\xv dE
\rho_{XE}(\xv,E,0)\,
\omega_F(\xv',E'|\xv,E), \nonumber\\
\label{PF-factor1}
\ea
where the reduced transition probability is defined as
\ba
&& \omega_F(\xv',E'|\xv,E) = \int d\yv d\yv' \label{omegaF-def} \\
&&\quad \frac{\delta\left(E-H_{XY}(\xv,\yv,0)\right)}{\Omega_Y(\xv,E,0)}
\delta\left(\xv'-\xv'(\xv,\yv)\right)
\nonumber\\
&&\quad
\times\delta\left(\yv'-\yv'(\xv,\yv)\right)
\delta\left(E'-H_{XY}(\xv',\yv',\tau)\right) \nonumber
\ea
with $(\xv_\tau,\yv_\tau)$ denoting the Hamiltonian evolution of $(\xv,\yv)$ from $t=0$ to $t=\tau$.
The reverse process is defined by the time-reversed protocol, with initial distribution 
\ba
\rho^R_{XE}(\xv^*,E,0)
=
\rho_{XE}(\xv,E,\tau).
\label{Crooks_choice_main}
\ea
Here $\xv^*$ denotes $\xv$ with all momenta reversed (likewise $\yv^*$ below). The corresponding reduced transition probability is
\ba
&&\omega_R (\xv^*,E|\xv'^*,E') = \int d\yv'^* d\yv^* \label{omegaR-def}\\
&&\quad
\frac{ \delta\left(E'-H^R_{XY}(\xv'^*,\yv'^*,0)\right) }{\Omega_Y^R(\xv'^*,E',0)}
\delta\left(\xv^*-\xv^{R}(\xv'^*,\yv'^*)\right)
\nonumber\\
&&\quad\times
\delta\left(\yv^*-\yv^{R}(\xv'^*,\yv'^*)\right)
\delta\left(E-H^R_{XY}(\xv^*,\yv^*,\tau)\right). \nonumber
\ea
where the reverse dynamics is generated by the Hamiltonian
$H^R_{XY}$ defined through
\ba
H^R_{XY}(\xv^*,\yv^*,t)
=
H_{XY}(\xv,\yv,\tau-t),
\ea
corresponding to the time-reversed protocol.

A key geometric identity, derived in Appendix~\ref{app:FT}, is
\ba
\frac{\omega_F(\xv_\tau,E_\tau|\xv_0,E_0)}
{\omega_R(\xv_0^*,E_0|\xv_\tau^*,E_\tau)}
=
\frac{\Omega_Y(\xv_\tau,E_\tau,\tau)}
{\Omega_Y(\xv_0,E_0,0)},
\label{state_ratio_Omega}
\ea
which follows from the one-to-one, measure-preserving nature of Hamiltonian dynamics together with the invariance of the Hamiltonian under momentum reversal.

Combining \eqref{state_ratio_Omega} with \eqref{Crooks_choice_main} gives
\ba
\log\frac{\mathcal P_F}{\mathcal P_R}
= s^{\rm md}(\xv_\tau,E_\tau,\tau)
- s^{\rm md}(\xv_0,E_0,0)
\equiv
\Delta s^{\rm md}.\nonumber\\
\ea
Averaging over the forward process in Eq.~(\ref{rho-tilde-main}), as propagated by the forward Hamiltonian dynamics, immediately yields the integral fluctuation theorem
\ba
\left\langle e^{-\Delta s^{\rm md}}\right\rangle_F
= 1.
\label{IFT-MDE}
\ea
Furthermore, grouping trajectories according to their entropy production gives the detailed fluctuation theorem
\ba
\frac{P_F(\Delta s)}{P_R(-\Delta s)}
= e^{\Delta s}.
\label{DFT-MDE}
\ea
Here the subscript $F$ denotes averaging over the forward-process ensemble. These fluctuation relations therefore follow directly from deterministic Hamiltonian dynamics. No stochastic assumptions are required: the apparent stochasticity arises solely from the reduced description in $(\xv,E)$ and the multiplicity of hidden microstates encoded by $\Omega_Y$.

Finally, while $s^{\rm md}$ is always well defined, its interpretation as an instantaneous thermodynamic entropy requires conditional equilibration of the bath variables $\yv$ at fixed $(\xv,E)$, as realized by the distribution $\tilde \rho_{XY}$. This distribution is the least structured microscopic realization consistent with a given reduced state $\rho_{XE}$: the unobserved Y variables carry no information beyond that encoded in $\Omega_Y(\xv,E)$.
In our proof of the fluctuation theorem \eqref{IFT-MDE} this conditional equilibration is assumed only for the initial/final states of the forward/reverse processes; no formal assumption is made about equilibration at intermediate times along the trajectory.

Our derivation of \eqref{IFT-MDE} is closely related to previous derivations of fluctuation relations in stochastic thermodynamics~\cite{Broeck2009,Evans1993,Gallavotti1995,Jarzynski1997,Crooks1999,Seifert2005,TalknerHanggi2007,Jarzynski2000}. As emphasized in Ref.~\cite{Broeck2009}, there exist infinitely many fluctuation theorems, corresponding to different choices of physical variables, path probabilities, and involutions. Many standard formulations arise either in stochastic settings, such as Langevin or Markov dynamics~\cite{Seifert2005,Crooks1999}, or from deterministic but non-Hamiltonian dynamics describing thermostatted systems~\cite{Evans1993,Gallavotti1995}. 

Of particular relevance to our work are formulations based on exact Hamiltonian dynamics~\cite{Jarzynski1997,Jarzynski2000}, in which the structure of the fluctuation theorem follows from Liouville's theorem together with time-reversal symmetry and an appropriate choice of forward and reverse ensembles. The closest antecedent to the present construction is the work of Jarzynski~\cite{Jarzynski2000}, who derived a detailed fluctuation relation for Hamiltonian systems coupled to heat reservoirs, without invoking additional dynamical assumptions beyond deterministic evolution, time-reversal symmetry, and canonical preparation of the reservoirs.

The present result shares this dynamical origin, but differs in the level of description: here the fluctuation relation is formulated for the reduced variables $(\xv,E)$, and the entropy change is given by the trajectory-level MDE, in which the conditional phase-space volume $\Omega_Y(\xv,E)$ appears explicitly. The only ensemble assumption is that, at the initial time (of both the forward and reversed protocol), the unmonitored Y variables are conditionally equilibrated in the sense defined above. With this choice, irreversibility is expressed directly in terms of the reduction from $(\xv,\yv)$ to $(\xv,E)$ and the associated multiplicity of unobserved microstates, rather than by invoking an externally imposed thermodynamic ensemble.

\section{Heat and thermodynamics from MDE}
\label{sec:thermo}

In general we can expect the total XY Hamiltonian to take the form
\ba
H_{XY} = H^0_X(\xv)+H^0_Y(\yv)+H^0_{I}(\xv,\yv)
\ea
where each term may depend also on time. The results presented above in the MDE framework are constructed directly from the full Hamiltonian $H_{XY}$ without any assumptions such as weak coupling or a large bath Y.

However, when we want to study heat and related physical quantities, thermodynamic considerations require partitioning the total energy into two parts, associated with the system X and the bath Y, so that heat can be defined as the energy transferred between them. In textbook statistical mechanics, it is assumed that the contribution from the interaction part $H^0_I$ can be directly ignored for the systems studied. This is justified when both the system X and the bath Y are large, and the interaction part $H^0_I$ is subextensive, scaling with the surface rather than the volume. However, this cannot be relied upon in more general situations, such as where the coupling is strong and long-range, or the surface contribution cannot be ignored because X is not a large system.

In the strong-coupling literature, this issue is resolved in the large-bath limit by introducing the Hamiltonian of mean force~\cite{jarzynski2004,Kirkwood1935,Campisi2009,Talkner2020,Burke2024,Miller2018}, which provides an effective two-part description without invoking a weak-coupling approximation.
In the MDE framework, the corresponding construction is extended to finite Hamiltonian systems through the CPV $\Omega_Y(\xv,E,t)$.
For a driven Hamiltonian $H_{XY}(\xv,\yv,t)$, this yields a state-dependent partition of the total energy into system and bath contributions, and thereby a consistent definition of heat as the part of the energy change of the X subsystem not attributable to external work interactions.
In contrast to the usual Hamiltonian of mean force construction, no large-bath assumption is required, and the construction applies to finite systems with arbitrary coupling strength.

Here we illustrate the main thermodynamic principles of the MDE framework and demonstrate their consistency by recovering key results of stochastic thermodynamics. A fuller development of thermodynamics within this framework will be presented in future work.

\subsection{Definition of internal energy and thermodynamics on the trajectory level}
\label{sec:EX}

We begin by introducing a reference description for the bath that is independent of the subsystem variables. The density of states of the bare bath is defined as
\ba
\Omega_Y^{(0)}(E) \equiv \int d\yv  \delta\left(E - H_Y^0(\yv)\right),
\label{CPV-0}
\ea
with entropy
\ba
S_Y^{(0)}(E) \equiv \log \Omega_Y^{(0)}(E).
\label{Sy0-def}
\ea
The corresponding inverse temperature is
\ba
\beta(E) = \frac{\partial S_Y^{(0)}(E)}{\partial E}, \quad T(E) = \beta(E)^{-1}.
\label{beta-def}
\ea
This temperature depends only on the bath and provides a natural thermodynamic grounding for what follows. Note that in the above definitions $E \equiv E_{XY} = E_Y$ because in the reference state, all the energy resides in the bath Y and none in the X subsystem. 

We now define an effective internal energy for the subsystem X by comparing the conditional phase-space volume with this bath reference,
\ba
E_X(\xv,E) = -T(E)\log \frac{\Omega_Y(\xv,E)}{\Omega_Y^{(0)}(E)}.
\label{EX-def-2}
\ea
This definition has the same structure as the Hamiltonian of mean force~\cite{Kirkwood1935,jarzynski2004,Talkner2020,Campisi2009}, but is formulated directly in terms of the phase-space geometry. It quantifies how the accessible phase-space volume of the bath is modified by the presence of the subsystem fixed at $\xv$, relative to the bare bath described by $\Omega_Y^{(0)}(E)$.

To understand its meaning, consider the case of a separable Hamiltonian (dependent on a parameter $\lambda$ that can be time-dependent),
\ba
H_{XY}(\xv,\yv,\lambda)=H_X^0(\xv,\lambda)+H_Y^0(\yv).
\ea
In this case the conditional phase-space volume becomes
\ba
\Omega_Y(\xv,E)=\Omega_Y^{(0)}\!\left(E-H_X^0(\xv,\lambda)\right),
\ea
so that $E_X(\xv,E)$ is expressed exactly in terms of a finite difference of the bath entropy. In the large-bath limit, where $S_Y^{(0)}(E)$ is smooth and can be linearized, one obtains
\ba
E_X(\xv,E)\approx H_X^0(\xv,\lambda),
\ea
showing that the present definition recovers the bare subsystem energy in this limit.

We next introduce thermodynamic quantities along a trajectory of the selected variables $\xv$. The bath energy is defined as
\ba
E_Y(\xv,E) = E - E_X(\xv,E),
\label{EY-def}
\ea
and the infinitesimal heat and work are defined by
\ba
\dbar \mathcal Q = -dE_Y, \quad \dbar \mathcal W = dE.
\label{heat-work-def}
\ea
With these definitions, the first law follows identically,
\ba
dE_X = \dbar \mathcal Q + \dbar \mathcal W.
\label{first-law-state-2}
\ea
Thus the decomposition induced by $\Omega_Y(\xv,E)$ directly yields a trajectory-level first-law structure.

We now relate this structure to the entropy of the bath. Rewriting~(\ref{EX-def-2}) gives
\ba
\log \Omega_Y(\xv,E) = -\beta(E) E_X(\xv,E) + \log \Omega_Y^{(0)}(E).
\label{Omega-EX-relation}
\ea
Taking the differential yields
\ba
d\log \Omega_Y = -\beta  dE_X - E_X  d\beta + \beta  dE.
\ea
Using the first law, we obtain
\ba
d\log \Omega_Y = -\beta  \dbar \mathcal Q - E_X  d\beta.
\label{dlogOmega-final}
\ea
This relation shows that changes in the conditional phase-space volume are directly governed by heat exchange with the bath. The first term has the standard thermodynamic form relating entropy change to heat, while the second term arises from the dependence of the temperature on the total energy and represents a finite-bath correction. It vanishes when the total energy is fixed, or in the large-bath limit where $\beta(E)$ is effectively constant.

Using Eq.~(\ref{dlogOmega-final}), the relation \eqref{state_ratio_Omega} for an infinitesimal transition from $(\xv,E)$ to $(\xv+d\xv,E+dE)$ in the reduced state space becomes
\ba
&&\log\frac{\omega_F(\xv_0+d\xv,E_0+dE |\xv_0,E_0)}
{\omega_R(\xv_0^*,E_0|\xv^*_0+d\xv^*,E_0+dE)} \nonumber\\
&&= -\beta  \dbar \mathcal Q - E_X  d\beta.
\label{local-balance-general}
\ea
When $d\beta$ is negligible, this reduces to the standard local detailed balance relation, from which a variety of fluctuation relations can be derived~\cite{Crooks1999,Seifert2005,ding2022}. Thus, in approaching the isothermal limit, the familiar local thermodynamic balance relation is recovered directly from the CPV ratio derived from Hamiltonian microreversibility.

Combining this with the trajectory-level entropy defined in Sec.~\ref{sec:MDEft}, we obtain
\ba
\Delta s_{XY}^{\rm md} = \Delta s_X - \beta Q,
\label{traj-entropy-thermo}
\ea
in any regime where the temperature is effectively constant. This recovers the standard structure of thermodynamic entropy production~\cite{Crooks1999,Seifert2005,Seifert2012,ding2022}.

A key conceptual advantage of our MDE framework is that $E_X(\xv,E)$ is defined without reference to temperature. The partition of the total energy into contributions associated with the macroscopic variables X and the bath variables Y is determined directly by the conditional phase-space volume $\Omega_Y(\xv,E)$, which depends only on the underlying Hamiltonian description. Temperature does not enter this construction, but instead emerges subsequently, for example through derivatives of the entropy with respect to energy in the large-bath limit. In this sense, the thermodynamic structure is not imposed through equilibrium ensembles, but arises from the geometry of the reduced phase space. Moreover, this structure exists {\em a priori}, no matter the relative sizes of the X and Y subsystems, and therefore allows one to address cases that are very far from the `large bath' limit, as demonstrated with examples (albeit with constant total energy and hence no work interactions) in Paper I.

\subsection{Thermodynamics on the ensemble level}

The thermodynamic quantities introduced above depend only on the reduced variables $(\xv,E)$. In particular, $E_X(\xv,E)$, $E_Y(\xv,E)$, $\dbar\mathcal Q$, and $\dbar\mathcal W$ are independent of the microscopic coordinates $\yv$. Their ensemble-level counterparts can therefore be obtained directly from the reduced distribution $\rho_{XE}(\xv,E,t)$. We define 
\ba
U_X(t)=\int d\xv dE\,\rho_{XE}(\xv,E,t)E_X(\xv,E),
\label{UX-def}
\ea
and (with $E = E_X+E_Y$)
\ba
\bar E(t)=\int d\xv dE\,\rho_{XE}(\xv,E,t)E.
\label{Ebar-def}
\ea
Since the trajectory-level thermodynamic differentials are
\ba
\dbar\mathcal W=dE,\qquad \dbar\mathcal Q=dE_X-dE,
\ea
their ensemble averages satisfy
\ba
\dbar W=d\bar E,\qquad \dbar Q=dU_X-d\bar E.
\label{QWbar-def}
\ea
Hence
\ba
dU_X=\dbar Q+\dbar W.
\label{first-law-ensemble}
\ea
Thus the first-law structure carries over directly to the ensemble level. At this stage, however, the thermodynamic state is characterized by the joint distribution $\rho_{XE}(\xv,E,t)$ rather than by $\rho_X(\xv,t)$ alone, because the effective subsystem energy $E_X(\xv,E)$ depends explicitly on the total energy.

\vspace{3mm}

\subsection{Thermodynamic limit}
\label{sec:thermodynamic-limit}
We now show that the generalized subsystem energy defined from the CPV in Eq.~(\ref{EX-def-2}) reduces to the conventional Hamiltonian of mean force (HMF)~\cite{jarzynski2004,Campisi2009,Talkner2020,Strasberg2021} when the unresolved environment becomes thermodynamically large. The HMF serves physically as an effective internal energy for $X$ in strongly coupled systems when the bath is large~\cite{ding2022strong}. 

Let $N$ denote the size of the environment $Y$, while the subsystem $X$ is kept finite. We consider the limit $N\to\infty$ at fixed environmental energy density $E/N$. We assume that
\ba
H_{XY}(\xv,\yv)-H_Y^0(\yv)=O(1)
\ea
as $N\to\infty$, so that fixing the subsystem state $\xv$ produces only a finite perturbation of the environmental Hamiltonian. In the following, we use the symbol $\simeq$ to denote equivalence in this limit.

Using the Fourier representation of the delta function, we write the two CPVs as
\ba
\Omega_Y^{(0)}(E)
&=&
\frac{1}{2\pi}
\int_{-\infty}^{\infty}d\eta\,
e^{\Psi_0(\eta;E)},
\nonumber\\
\Omega_Y(\xv,E)
&=&
\frac{1}{2\pi}
\int_{-\infty}^{\infty}d\eta\,
e^{\Psi_0(\eta;E)+\Delta\Phi_{\xv}(\eta)},
\label{CPV-Fourier-pair}
\ea
where the exponents are
\ba
\Psi_0(\eta;E)
&=& i\eta E+ \log\int d\yv\,
e^{-i\eta H_Y^0(\yv)},
\nonumber\\
\Delta\Phi_{\xv}(\eta)
&=& \log
\frac{ \int d\yv\,
e^{-i\eta H_{XY}(\xv,\yv)} }
{ \int d\yv\,
e^{-i\eta H_Y^0(\yv)} }.
\label{Fourier-exponents}
\ea
Under the usual thermodynamic scaling, the exponent $\Psi_0$ is $O(N)$, whereas the finite perturbation $\Delta\Phi_{\xv}$ and its derivatives remain $O(1)$.

Denote the saddle point of $\Omega_Y^{(0)}$ by
\ba
\eta_0^*(E)=-i\beta_{\rm sp}(E).
\ea
The finite perturbation $\Delta\Phi_{\xv}$ shifts this saddle by only $O(N^{-1})$. The resulting correction to the exponent and the logarithm of the ratio of the Gaussian prefactors are therefore both $O(N^{-1})$. Under the usual regular saddle-point conditions, this gives
\ba
\log
\frac{\Omega_Y(\xv,E)}
{\Omega_Y^{(0)}(E)}
=
\Delta\Phi_{\xv}\!\left(-i\beta_{\rm sp}(E)\right)
+O(N^{-1}).
\label{CPV-ratio-saddle}
\ea
The saddle-point inverse temperature satisfies
\ba
\beta_{\rm sp}(E)
&=&
\partial_E\log\Omega_Y^{(0)}(E)+O(N^{-1})
\nonumber\\
&=&
\beta(E)+O(N^{-1}),
\label{saddle-microcanonical-temperature}
\ea
where $\beta(E)$ is the exact microcanonical inverse temperature defined in Eq.~(\ref{beta-def}). Since $\Delta\Phi_{\xv}(-i\beta)$ varies smoothly with $\beta$, replacing $\beta_{\rm sp}(E)$ by $\beta(E)$ introduces only a subdominant correction. We therefore obtain
\ba
\log
\frac{\Omega_Y(\xv,E)}
{\Omega_Y^{(0)}(E)}
\simeq
\log
\frac{
\int d\yv\,
e^{-\beta(E)H_{XY}(\xv,\yv)}
}{
\int d\yv\,
e^{-\beta(E)H_Y^0(\yv)}
}.
\label{CPV-ratio-thermodynamic-limit}
\ea
Combining this result with Eq.~(\ref{EX-def-2}) gives
\ba
E_X(\xv,E)
&\simeq&
H_X^{\rm HMF}\!\left(\xv;\beta(E)\right)
\label{HMF-definition-thermo-limit}\\
&\equiv&
-\frac{1}{\beta(E)}
\log
\frac{
\int d\yv\,
e^{-\beta(E)H_{XY}(\xv,\yv)}
}{
\int d\yv\,
e^{-\beta(E)H_Y^0(\yv)}
}.\nonumber
\ea
Thus, the HMF emerges directly from the CPV-defined subsystem energy in the thermodynamic limit, rather than being introduced as an independent definition. Its inverse temperature is fixed by the microcanonical reference environment.

Finally, suppose that the energy marginal is concentrated around a macroscopic value $E^*$ with a width of $O(\sqrt{N})$ or smaller. For a smooth environmental equation of state,
\ba
\beta(E)=\beta^*+O(N^{-1/2}),
\qquad
\beta^*\equiv\beta(E^*),
\label{beta-concentration}
\ea
throughout the region containing asymptotically all of the probability. The generalized internal energy then reduces to
\ba
U_X &=&
\int d\xv dE
\rho_{XE}(\xv,E,t)E_X(\xv,E)
\nonumber\\
&\simeq& \int d\xv
\rho_X(\xv,t) H_X^{\rm HMF}(\xv;\beta^*).
\label{UX-rhoX}
\ea
Under the same thermodynamic-limit assumptions, Eq.~(\ref{mde-shell-decomposition}) gives
\ba
\Delta S_{XY}^{\rm md}
\simeq
\Delta S_X^G-\beta^*Q,
\label{dSmd-ensemble}
\ea
which recovers the familiar expression for the total entropy production.

\vspace{5mm}
\section{Example: A driven particle gas with exactly calculable entropy production}
\label{sec:example}

A central implication of the present framework is that a thermodynamic limit is not required for a reduced thermodynamic description. The relevant physical requirement is instead that the unmonitored degrees of freedom mix sufficiently rapidly relative to the selected variables $\xv\in$ X, although such a separation is inevitably imperfect in real, finite systems. Here we study the example of driven six-particle gas (in two spatial dimensions) as a finite-system example in which no thermodynamic limit is available. We demonstrate that the endpoint MDE construction remains operational under these conditions and yields forward and reverse statistics in excellent agreement with the detailed and integral fluctuation theorems.

To illustrate this, we now give an explicit example in which the CPV in both the initial and final states can be found analytically. This is done by choosing a protocol in which those states are both noninteracting, while the intermediate Hamiltonian dynamics may involve any chosen interactions and/or external driving. This example provides a direct test of the trajectory-level fluctuation relation and illustrates that the entropy production is determined entirely by the endpoint CPVs, without requiring an explicit evaluation of the CPV during the intermediate protocol.

Consider $N$ identical particles of mass $m$ in a two-dimensional rectangular box of side lengths $L_x$ and $L_y$. The endpoint Hamiltonians both correspond to a gas of noninteracting point particles, 
\ba
H_0(\Gamma)=H_\tau(\Gamma)
= \frac{|{\mathbf p}|^2}{2m},
\label{endpoint-point-H}
\ea
where $\mathbf p$ denotes the momentum of the total system. We choose the reduced variables to be the center-of-position variables
\ba
X=(\bar x,\bar y),
\qquad
\bar x=\frac{1}{N}\sum_{i=1}^N x_i,
\qquad
\bar y=\frac{1}{N}\sum_{i=1}^N y_i.
\label{gas-X-def}
\ea
The CPV in both the initial ($t=0$) and the final ($t=\tau$) states obeys (with respective energy values $E_0,E_\tau$)
\ba
\Omega_Y(\bar x,\bar y,E)
&=&
\frac{1}{N!} \int d\Gamma
\delta\left(E-\sum_{i=1}^N\frac{p_{xi}^2+p_{yi}^2}{2m}\right)
\label{gas-CPV-def}\\
&&\times
\delta\left(\bar x-\frac{1}{N}\sum_{i=1}^N x_i\right)
\delta\left(\bar y-\frac{1}{N}\sum_{i=1}^N y_i\right)
\nonumber
\ea
where $d\Gamma = d\xv d\yv d\pv_x d \pv_y$ denotes the integration volume element of the full phase-space.
Here $Y$ denotes all remaining degrees of freedom after fixing $(\bar x,\bar y,E)$. The factor $1/N!$ is the usual Gibbs factor whose role in the MDE setting is fully discussed in Paper I. This constant contribution to $\Omega_Y$ plays no part in the comparisons that follow.

Because the initial and final Hamiltonians contains no interactions, the position and momentum integrations separate at both endpoints of any trajectory. The momentum-shell contribution is
\ba
\Omega_p(E)
=
\frac{(2\pi m)^N}{\Gamma(N)}
E^{N-1}.
\label{gas-momentum-shell}
\ea
The position integrals reduce to
\ba
\Omega_x(\bar x)
=
N L_x^{N-1}
f_N\left(\frac{N\bar x}{L_x}\right),
\ea
\ba
\Omega_y(\bar y)
=
N L_y^{N-1}
f_N\left(\frac{N\bar y}{L_y}\right).
\ea
Here $f_N(s)$ denotes the Irwin--Hall density, {\em i.e.}, the probability density for the sum of $N$ independent random variables uniformly distributed on $[0,1]$:
\ba
f_N(s)
&=&
\frac{1}{(N-1)!}
\sum_{k=0}^{\lfloor s\rfloor}
(-1)^k
{N\choose k}
(s-k)^{N-1},
\label{Irwin-Hall-density}
\ea
Combining these factors gives
\ba
\Omega_Y(\bar x,\bar y,E)
&=&
\frac{N^2L_x^{N-1}L_y^{N-1}}{N!}
f_N\left(\frac{N\bar x}{L_x}\right)
f_N\left(\frac{N\bar y}{L_y}\right) \nonumber\\
&&\frac{(2\pi m)^N}{\Gamma(N)}
E^{N-1}.
\label{gas-endpoint-CPV}
\ea

For a forward trajectory from $(X_0,E_0)$ to $(X_\tau,E_\tau)$, the trajectory-level change in MDE between initial and final states is
\ba
\Delta s^{\rm md}
= \log \frac{\rho_{XE}^F(X_0,E_0,0)}
{\rho_{XE}^F(X_\tau,E_\tau,\tau)}
+ \log \frac{\Omega_Y(X_\tau,E_\tau,\tau)}
{\Omega_Y(X_0,E_0,0)}
\label{gas-MDE-change-general}
\ea
which does not depend on the chosen Hamiltonian at intermediate times.
With a time-independent box geometry $L_{x,y}$, substituting Eq.~(\ref{gas-endpoint-CPV}) gives
\ba
\Delta s^{\rm md}
=
\log
\frac{\rho_{XE}^F(X_0,E_0,0)}
{\rho_{XE}^F(X_\tau,E_\tau,\tau)}
+(N-1)\log\frac{E_\tau}{E_0}
\nonumber\\
+\log
\frac{
f_N\left(N\bar x_\tau/L_x\right)
f_N\left(N\bar y_\tau/L_y\right)
}{
f_N\left(N\bar x_0/L_x\right)
f_N\left(N\bar y_0/L_y\right)
}.
\label{gas-MDE-change-explicit}
\ea
The explicit construction for the initial and final MDEs in \eqref{gas-endpoint-CPV} thus determines the trajectory-level entropy production in terms of the endpoint values of $(X,E)$ on that trajectory. The intermediate protocol encoded in the time-dependent Hamiltonian serves only to deliver the system from $(X_0,E_0)$ to $(X_\tau,E_\tau)$.

To test numerically the fluctuation relation in Eq.~(\ref{DFT-MDE}), we retain Eq.~(\ref{endpoint-point-H}) at both endpoints but use non-trivial dynamics during the protocol itself. Specifically, a bounded repulsive interaction between the particles is switched on and off during the interval $0<t<\tau$ as
\ba
U_{\rm rep}(\Gamma,t)
=
\sum_{i<j}
\epsilon(t)
\exp\left[
-\frac{|\mathbf r_i-\mathbf r_j|^2}{2\sigma^2}
\right],
\ea
with
\ba
\epsilon(t)
=
\epsilon_{\max}
\sin^2\left(\frac{\pi t}{\tau}\right).
\label{gas-soft-core}
\ea
In addition, a spatially uniform driving force in the $x$ direction is applied by introducing a drive potential
\ba
U_{\rm drive}(\Gamma,t)
=
-F_0
\sin\left(\frac{\pi t}{\tau}\right)
\sum_{i=1}^N x_i.
\label{gas-driving-potential}
\ea
These interactions modify the trajectory statistics and the final distribution of $(X,E)$, but do not invalidate the exact calculation of initial and final MDEs via \eqref{gas-endpoint-CPV}.

Next we specify the initial conditions by choosing a distribution on $(\bar x,\bar y)$ as a product of truncated Gaussians (set to zero outside the box) with a truncated Gamma distribution in $E$. Initial microstates are sampled uniformly from the corresponding CPV shell at each $\bar x, \bar y, E$. The reverse ensemble is constructed from the empirical forward final reduced distribution using the standard time-reversal procedure, and the final reduced density $\rho_{XE}^F(X,E,\tau)$ is estimated from the forward trajectories.

\begin{figure}[t]
\centering
\includegraphics[width=0.45\textwidth]{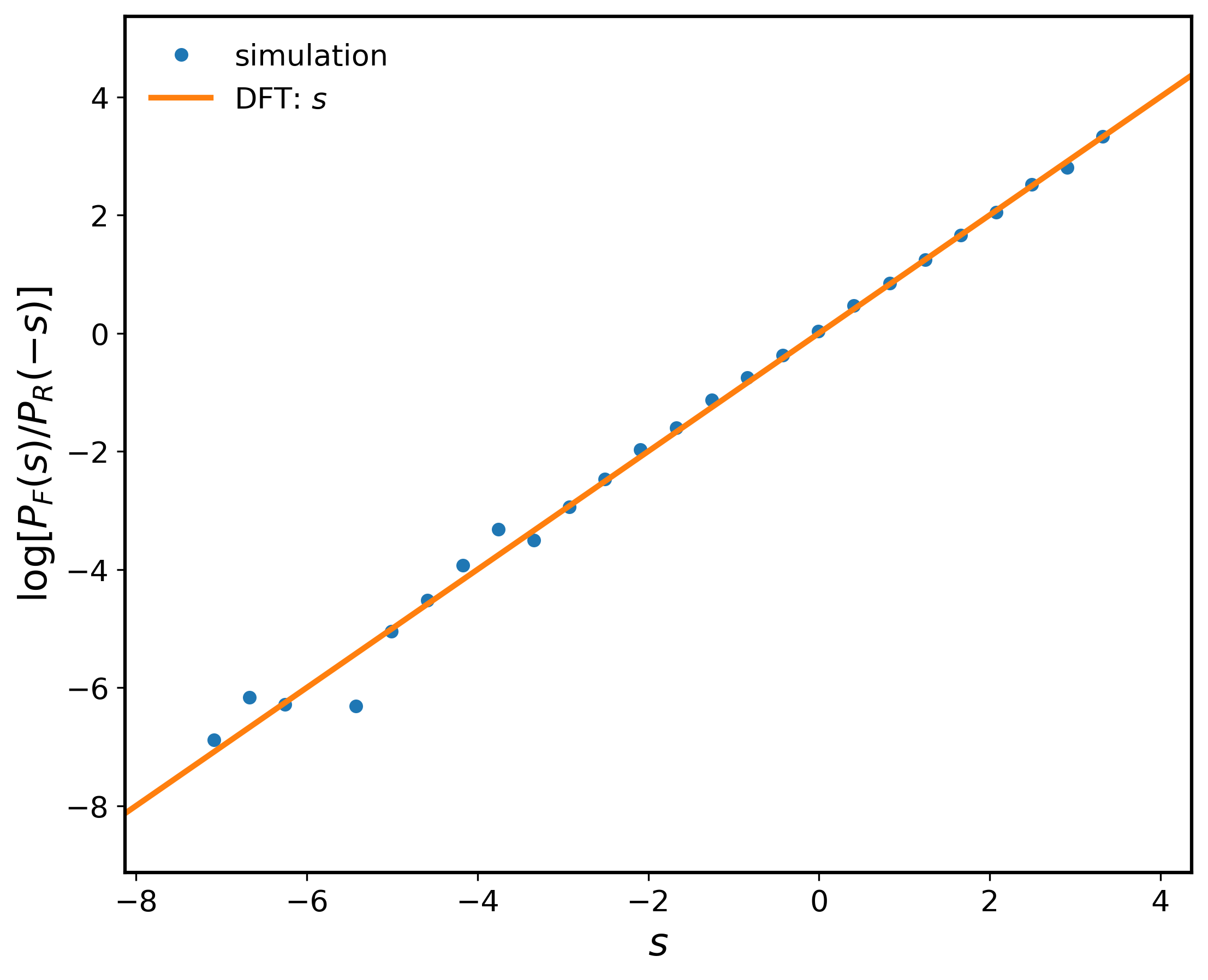}
\caption{
Numerical verification of the endpoint-CPV detailed fluctuation theorem for the driven particle gas. The blue symbols show the measured value of $\log[P_F(s)/P_R(-s)]$, where $s=\Delta s^{\rm md}$ is computed from Eq.~(\ref{gas-MDE-change-explicit}). The orange line shows the theoretical prediction $\log[P_F(s)/P_R(-s)]=s$. The agreement confirms that the detailed fluctuation theorem is controlled by the endpoint CPVs, while the intermediate interaction and driving only affect the trajectory statistics.
}
\label{fig:endpoint-cpv-dft}
\end{figure}

In the simulations we used $N=6$, $m=1$, $L_x=L_y=1$, $\tau=5$, $F_0=0.25$, $\epsilon_{\max}=0.08$, and $\sigma=0.10$. The forward and reverse ensembles each contained $4.5\times10^4$ independently sampled trajectories. The resulting estimate of
\ba
\log\frac{P_F(\Delta s)}
{P_R(-\Delta s)} \label{eq:ST_EPR}
\ea
agrees with the theoretical prediction $\Delta s$, as shown in Fig.~\ref{fig:endpoint-cpv-dft}. This confirms the detailed fluctuation theorem and demonstrates that the entropy production is completely determined by the endpoint CPVs. 

Normally, the calculation of entropy production via \eqref{eq:ST_EPR} is generally based on a formulation of stochastic thermodynamics in which the Y variables are assumed to be very large in number so that their influence can be replaced by a Langevin dynamics or similar construction. Yet, remarkably, here we see the result quantitatively verified for a system of just $6$ particles (so that Y contains a total of just 22 coordinates/momenta). Thus, introduction of the MDE formalism extends the domain of stochastic thermodynamics beyond the usual case of a small number of degrees of freedom interacting with a large bath, to the case when the bath is likewise small.

\vspace{5mm}
\section{Conclusion}
\label{sec:conclusion}

We have developed a general statistical framework for Hamiltonian systems based on a transform from the full phase space $(\xv,\yv)$ to the variables $(\xv,E)$, where $\xv$ denotes the variables describing macroscopic thermal properties, and $E$ is the total energy. This transform retains the measure structure induced by Hamiltonian dynamics while eliminating unobserved microscopic details in $\yv$, and thus constitutes a non-destructive coarse-graining, in the sense that the reduced dynamics of $\rho_{XE}$ is obtained exactly from the underlying Hamiltonian evolution without modification. At the same time, this reduction induces a probabilistic description: the deterministic dynamics in the full phase space gives rise to probabilistic evolution at the level of $(\xv,E)$ through the multiplicity of microstates encoded in the conditional phase-space volume. The central object is the CPV $\Omega_Y(\xv,E)$, which counts the unobserved microstates compatible with a given configuration and energy and provides a geometric characterization of equilibrium entropy.

Building on Ref.~\cite{Ding2026}, where monotonicity of the MDE was established under broad conditions for relaxation on a fixed energy shell, we have generalized the construction to arbitrary phase-space distributions. This allows work interactions to be included via time-dependent Hamiltonian protocols. The resulting entropy admits a decomposition Eq.~\eqref{mde-kl-form} into a structural (equilibrium) contribution determined by $\Omega_Y$ and a Kullback-Leibler term that quantifies deviations from conditional equilibrium. Our work shows that irreversibility arises from a choice of description that disregards unobserved degrees of freedom, even while the underlying Hamiltonian dynamics remains exactly reversible.

In addition, we have shown that the same framework naturally induces a thermodynamic structure. The conditional phase-space volume $\Omega_Y(\xv,E)$ determines a decomposition of the total energy into system and bath contributions, from which heat, work, and entropy balance relations follow directly at the trajectory level. In the large-bath limit, this structure reduces to standard stochastic thermodynamics, demonstrating that conventional, ensemble-based thermodynamics emerges as a limiting description of the MDE framework. However, we have strikingly shown that by defining the MDE without assuming a large bath, the domain of stochastic thermodynamics can be extended to `baths' of only a few particles ($N=6$, in the example discussed in Sec.~\ref{sec:example}).

Taken together with the findings of Paper I, these results provide a unified framework in which entropy and thermodynamic structure emerge from a single quantity, $\Omega_Y(\xv,E)$. Defined on the reduced state space $(\xv,E)$ associated with a time-dependent Hamiltonian $H_{XY}(t)$, the CPV provides a common geometric foundation for both equilibrium and nonequilibrium phenomena.

Within our MDE framework, thermodynamic behavior is obtained directly from the Hamiltonian dynamics and phase-space geometry of the full XY system, without introducing stochastic assumptions, idealized reservoirs, or modifications of the microscopic equations of motion. The central requirement for the emergence of thermodynamics is the autonomy condition, namely that the evolution of the reduced state density $\rho_{XE}$ is independent of the detailed bath dynamics and depends only on the information retained in $(\xv,E)$. Under this condition, the MDE framework provides a microscopic route from reversible Hamiltonian dynamics to emergent, classical thermodynamic irreversibility~\cite{Ding2026}. Moreover, as shown in the present paper, this route also encompasses (alongside work interactions) the main precepts of stochastic thermodynamics, and extends these to systems where the heat bath is, from any conventional thermodynamic perspective, unreasonably small.

\bibliographystyle{unsrt}
\bibliography{reference}

\appendix

\begin{widetext}
\section{Trajectory-level MDE and fluctuation relation}
\label{app:FT}

We define the trajectory-level Microscopic Dynamical Entropy (MDE) as
\ba
s^{\rm md}(\xv,E,t)
=-\log \rho_{XE}(\xv,E,t)
+\log \Omega_Y(\xv,E,t).
\label{smd-def}
\ea
We denote by $\xv^*$ and $\yv^*$ the time-reversed phase-space points, obtained by reversing the sign of all momenta while keeping coordinates unchanged. We assume Hamiltonian reversibility as
\ba
H_{XY}(\xv^*,\yv^*,t)=H_{XY}(\xv,\yv,t),
\ea
so that we have
\ba
\Omega_Y(\xv^*,E,t)=\Omega_Y(\xv,E,t).
\label{symmetry}
\ea
We consider a transition process from $(\xv,\yv)$ at $t=0$ to $(\xv',\yv')$ at $t=\tau$ under Hamiltonian evolution. The reversed dynamics is generated by the Hamiltonian $H^R_{XY}$ under the time-reversed protocol, satisfying
\ba
H^R_{XY}(\xv^*,\yv^*,t)=H_{XY}(\xv,\yv,\tau-t).
\label{H-reverse}
\ea
The corresponding reversed transition process is from $(\xv'^*,\yv'^*)$ to $(\xv^*,\yv^*)$. Microreversibility implies the trajectory correspondence
\ba
(\xv',\yv')=(\xv'(\xv,\yv),\yv'(\xv,\yv))
\Longleftrightarrow
(\xv^*,\yv^*)=(\xv^{R}(\xv'^*,\yv'^*),\yv^{R}(\xv'^*,\yv'^*)).
\label{microrev}
\ea
We define the reverse conditional phase volume as
\ba
\Omega_Y^R(\xv,E,t)
=\int d\yv\,
\delta\left(E-H^R_{XY}(\xv,\yv,t)\right).
\label{OmegaR}
\ea
Using Eq.~(\ref{H-reverse}) and relabeling $\yv^*$ as $\yv$, one obtains for all $t$
\ba
\Omega_Y^R(\xv^*,E,t)=\Omega_Y(\xv,E,\tau-t).
\ea
In the full phase space $(\xv,\yv)$, the dynamics is deterministic. But when we transform to the variables $(\xv,E)$, the description becomes probabilistic since the projection from $(\xv,\yv)$ to $(\xv,E)$ is many-to-one.
We therefore construct a corresponding phase-space distribution by distributing probability uniformly over the unobserved variables $\yv$ at fixed $(\xv,E)$ as
\ba
\tilde\rho_{XY}(\xv,\yv,0)
=\frac{\rho_{XE}(\xv,H_{XY}(\xv,\yv,0),0)}
{\Omega_Y(\xv,H_{XY}(\xv,\yv,0),0)}.
\label{rho-tilde}
\ea
The probability density to reach $(\xv',E')$ is
\ba
\mathcal P_F(\xv',E')
&=&
\int d\xv\, d\yv\;
\tilde\rho_{XY}(\xv,\yv,0)
\nonumber\\
&&\times
\delta\left(\xv'-\xv'(\xv,\yv)\right)
\delta\left(E'-H_{XY}\left(\xv'(\xv,\yv),\yv'(\xv,\yv),\tau\right)\right).
\label{PF-origin}
\ea
We define $\rho_{XE}(\xv',E',\tau)$ as the reduced distribution obtained by evolving the reconstructed initial ensemble in Eq.~(\ref{rho-tilde}) under the forward Hamiltonian dynamics. The probability density in Eq.~(\ref{PF-origin}) is therefore
\ba
\mathcal P_F(\xv',E')
=
\rho_{XE}(\xv',E',\tau).
\ea
Substituting Eq.~(\ref{rho-tilde}) and reorganizing the integral using $E=H_{XY}(\xv,\yv,0)$ gives
\ba
\mathcal P_F(\xv',E')
&=&
\int d\xv\, dE\;
\rho_{XE}(\xv,E,0)\,
\omega_F(\xv',E'|\xv,E),
\label{PF-factor2}
\ea
where $\omega_F(\xv',E'|\xv,E)$ is given by
\ba
\omega_F(\xv',E'|\xv,E)
&=&
\frac{1}{\Omega_Y(\xv,E,0)}
\int d\yv\, d\yv'
\delta\left(E-H_{XY}(\xv,\yv,0)\right)
\delta\left(\xv'-\xv'(\xv,\yv)\right)
\nonumber\\
&&\times
\delta\left(\yv'-\yv'(\xv,\yv)\right)
\delta\left(E'-H_{XY}(\xv',\yv',\tau)\right).
\label{omegaF}
\ea
We can similarly define the reverse transition weight as
\ba
\omega_R(\xv^*,E|\xv'^*,E')
&=&
\frac{1}{\Omega_Y^R(\xv'^*,E',0)}
\int d\yv'^*\, d\yv^*
\delta\left(E'-H^R_{XY}(\xv'^*,\yv'^*,0)\right)
\delta\left(\xv^*-\xv^{R}(\xv'^*,\yv'^*)\right)
\nonumber\\
&&\times
\delta\left(\yv^*-\yv^{R}(\xv'^*,\yv'^*)\right)
\delta\left(E-H^R_{XY}(\xv^*,\yv^*,\tau)\right).
\label{omegaR}
\ea
The delta functions in Eq.~(\ref{omegaF}) restrict $(\xv,\yv)$ and $(\xv',\yv')$ to lie on the same Hamiltonian trajectory. Microreversibility establishes a one-to-one correspondence between this forward trajectory and the reversed trajectory from $(\xv'^*,\yv'^*)$ to $(\xv^*,\yv^*)$. Moreover, both Hamiltonian evolution and momentum reversal preserve the phase-space measure: the Hamiltonian flow has unit Jacobian by Liouville's theorem, while momentum reversal has unit absolute Jacobian. Therefore, as distributions under the phase-space integrals,
\ba
\delta\left(\xv'-\xv'(\xv,\yv)\right)
\delta\left(\yv'-\yv'(\xv,\yv)\right)
=\delta\left(\xv^*-\xv^{R}(\xv'^*,\yv'^*)\right)
\delta\left(\yv^*-\yv^{R}(\xv'^*,\yv'^*)\right).
\label{delta-map}
\ea
Using Eq.~(\ref{H-reverse}), the energy-shell constraints transform as
\ba
\delta\left(E-H_{XY}(\xv,\yv,0)\right)
&=&\delta\left(E-H^R_{XY}(\xv^*,\yv^*,\tau)\right),\\
\delta\left(E'-H_{XY}(\xv',\yv',\tau)\right)
&=&\delta\left(E'-H^R_{XY}(\xv'^*,\yv'^*,0)\right).
\ea
We now relabel the dummy variables $(\yv,\yv')\to(\yv^*,\yv'^*)$, which is a one-to-one relabeling of integration variables. After this step, the integrand of $\omega_F$ becomes identical to that of $\omega_R$, so the numerators coincide.
From the definition of $\Omega_Y^R$ and Eq.~(\ref{H-reverse}), relabeling $\yv'^*$ as $\yv'$ gives
\ba
\Omega_Y^R(\xv'^*,E',0)=\Omega_Y(\xv',E',\tau).
\label{Omega-map}
\ea
Hence we obtain the relation
\ba
\frac{\omega_F(\xv',E'|\xv,E)}
{\omega_R(\xv^*,E|\xv'^*,E')}
=\frac{\Omega_Y(\xv',E',\tau)}{\Omega_Y(\xv,E,0)}.
\label{omega-ratio}
\ea

For a reduced endpoint transition
$\Gamma:(\xv_0,E_0)\to(\xv_\tau,E_\tau)$, define
\ba
\mathcal P_F[\Gamma]
=\rho_{XE}(\xv_0,E_0,0)
\omega_F(\xv_\tau,E_\tau|\xv_0,E_0),
\ea
and
\ba
\mathcal P_R[\Gamma^R]
=\rho^R_{XE}(\xv_\tau^*,E_\tau,0)
\omega_R(\xv_0^*,E_0|\xv_\tau^*,E_\tau).
\ea
This gives
\ba
\log\frac{\mathcal P_F[\Gamma]}{\mathcal P_R[\Gamma^R]}
=-\log \rho^R_{XE}(\xv_\tau^*,E_\tau,0)
+\log \rho_{XE}(\xv_0,E_0,0)
\nonumber\\
+\log \Omega_Y(\xv_\tau,E_\tau,\tau)
-\log \Omega_Y(\xv_0,E_0,0).
\ea
Choosing $\rho^R_{XE}(\xv^*,E,0)=\rho_{XE}(\xv,E,\tau)$ yields
\ba
\log\frac{\mathcal P_F[\Gamma]}{\mathcal P_R[\Gamma^R]}
=\Delta s^{\rm md}[\Gamma].
\label{traj-identity}
\ea
For the corresponding reversed transition
$\Gamma^R:(\xv_\tau^*,E_\tau)\to(\xv_0^*,E_0)$,
we define the reverse entropy-production functional as
\ba
\Delta s_R^{\rm md}[\Gamma^R]
\equiv
\log\frac{\mathcal P_R[\Gamma^R]}
{\mathcal P_F[\Gamma]}.
\label{reverse-smd-def}
\ea
Using the definitions of the forward and reverse transition probabilities,
this can be written explicitly as
\ba
\Delta s_R^{\rm md}[\Gamma^R]
&=&
\log\rho^R_{XE}(\xv_\tau^*,E_\tau,0)
-\log\rho_{XE}(\xv_0,E_0,0)
\nonumber\\
&&+\log\Omega_Y^R(\xv_0^*,E_0,\tau)
-\log\Omega_Y^R(\xv_\tau^*,E_\tau,0).
\label{reverse-smd-explicit}
\ea
Using
$\rho^R_{XE}(\xv_\tau^*,E_\tau,0)
=\rho_{XE}(\xv_\tau,E_\tau,\tau)$
and
$\Omega_Y^R(\xv^*,E,t)
=\Omega_Y(\xv,E,\tau-t)$,
we obtain
\ba
\Delta s_R^{\rm md}[\Gamma^R]
=-\Delta s^{\rm md}[\Gamma].
\label{smd-antisymmetry}
\ea
For the forward process,
$\Delta s^{\rm md}[\Gamma]$ coincides with the change in the
trajectory-level MDE between the two endpoints. By contrast,
$\Delta s_R^{\rm md}[\Gamma^R]$ is the reverse entropy-production
functional defined by exchanging the forward and reverse transition
probabilities. In general, it is not equal to the change in the MDE
evaluated using the actual time-dependent reduced distribution generated
by the independently evolving reverse ensemble. In particular, the
density associated with the final reverse endpoint in
Eq.~(\ref{reverse-smd-explicit}) is the forward initial density
$\rho_{XE}(\xv_0,E_0,0)$, rather than the actual reverse final density
$\rho^R_{XE}(\xv_0^*,E_0,\tau)$.

We define the probability distributions of the forward and reverse
entropy-production functionals as
\ba
P_F(\Delta s)
&=&
\int\mathcal D\Gamma\;
\mathcal P_F[\Gamma]\,
\delta\left(
\Delta s-\Delta s^{\rm md}[\Gamma]
\right),\\
P_R(\Delta s)
&=&
\int\mathcal D\Gamma^R\;
\mathcal P_R[\Gamma^R]\,
\delta\left(
\Delta s-\Delta s_R^{\rm md}[\Gamma^R]
\right).
\ea
Here $\mathcal D\Gamma$ denotes integration over the initial and final
reduced endpoints $\mathcal D\Gamma
=d\xv_0\,dE_0\,d\xv_\tau\,dE_\tau$ with the analogous definition for $\mathcal D\Gamma^R$.
Changing the integration variable from $\Gamma$ to $\Gamma^R$ and using
Eqs.~(\ref{traj-identity}) and (\ref{smd-antisymmetry}), we obtain
\ba
P_F(\Delta s)
&=&
\int\mathcal D\Gamma\;
\mathcal P_R[\Gamma^R]\,
e^{\Delta s^{\rm md}[\Gamma]}
\delta\left(
\Delta s-\Delta s^{\rm md}[\Gamma]
\right)
\nonumber\\
&=&
e^{\Delta s}
\int\mathcal D\Gamma^R\;
\mathcal P_R[\Gamma^R]\,
\delta\left(
-\Delta s-\Delta s_R^{\rm md}[\Gamma^R]
\right)
\nonumber\\
&=&
e^{\Delta s}P_R(-\Delta s).
\ea
Therefore, the detailed fluctuation theorem is
\ba
\frac{P_F(\Delta s)}
{P_R(-\Delta s)}
=e^{\Delta s}.
\label{DFT}
\ea
The corresponding integral fluctuation theorem follows directly from
Eq.~(\ref{traj-identity}):
\ba
\left\langle e^{-\Delta s^{\rm md}}\right\rangle_F
&=&
\int\mathcal D\Gamma\;
\mathcal P_F[\Gamma]\,
e^{-\Delta s^{\rm md}[\Gamma]}
= \int\mathcal D\Gamma^R\;
\mathcal P_R[\Gamma^R] =1.
\label{IFT}
\ea

\end{widetext}

\end{document}